\documentclass[10pt]{IEEEtran}
\nonstopmode
\IEEEoverridecommandlockouts

\usepackage{cite}
\usepackage{amsmath,amssymb,amsfonts}
\usepackage{algorithmic}
\usepackage{graphicx}
\usepackage{textcomp}
\usepackage{xcolor}

\usepackage{comment}

\usepackage{colortbl}
\usepackage{dhucs} 

\usepackage{booktabs} 
\usepackage{multirow}

\usepackage{makecell}
\usepackage{tabularx}
\usepackage{subcaption}
\usepackage{bm}

\usepackage{color} 
\usepackage{setspace} 
\usepackage{graphicx} 
\usepackage{makecell}
\usepackage{enumitem} 
\usepackage{nicematrix} 
\usepackage{threeparttable} 
\usepackage{dblfloatfix}    

\newcommand{\editorialcomment}[1]{\textcolor{blue}}
\newcommand{\delete}[1]{
}

\newcommand\refFigure[1]{Fig.~\ref{#1}}
\newcommand\refTable[1]{Table~\ref{#1}}

\newcommand\korean[1]{}
\newcommand\blind[1]{XXXX}

\usepackage{relsize}

\def\BibTeX{{\rm B\kern-.05em{\sc i\kern-.025em b}\kern-.08em
    T\kern-.1667em\lower.7ex\hbox{E}\kern-.125emX}}
\begin{document}

\title{Standalone FPGA-Based QAOA Emulator for Weighted-MaxCut on Embedded Devices
}

\author{\large
Seonghyun Choi, Kyeongwon Lee, Jae-Jin Lee, and Woojoo Lee 
}

\maketitle

\begin{abstract}
Quantum computing (QC) emulation is crucial for advancing QC applications, especially given the scalability constraints of current devices. 
FPGA-based designs offer an efficient and scalable alternative to traditional large-scale platforms, but most are tightly integrated with high-performance systems, limiting their use in mobile and edge environments. 
This study introduces a compact, standalone FPGA-based QC emulator designed for embedded systems, leveraging the Quantum Approximate Optimization Algorithm (QAOA) to solve the Weighted-MaxCut problem. 
By restructuring QAOA operations for hardware compatibility, the proposed design reduces time complexity from $\bm{O(N^2)}$ to $\bm{O(N)}$, where $\bm{N = 2^n}$ for $\bm{n}$ qubits. 
This reduction, coupled with a pipeline architecture, significantly minimizes resource consumption, enabling support for up to 9-qubits on mid-tier FPGAs—three times more than comparable designs. 
Additionally, the emulator achieved energy savings ranging from 1.53$\bm{\times}$ for 2-qubits to up to 852$\bm{\times}$ for 9-qubits compared to software-based QAOA implementations on embedded processors.
These results highlight the practical scalability and resource efficiency of the proposed design, providing a robust foundation for QC emulation in resource-constrained edge devices.


\end{abstract}

\section{Introduction}

Over recent decades, quantum computing (QC) has made remarkable strides, leveraging principles of superposition, entanglement, and parallelism to surpass classical computing in specific domains. 
These advancements have garnered significant attention from both academia~\cite{Wang:ACM22,Kim:Nature23, Kikuchi:npjQI23,Almudever:DATE24} and industry~\cite{Microsoft:Qubit, Google:Science20}, firmly establishing QC as a central focus of modern research. 
However, current QC devices face substantial scalability and error correction challenges, preventing them from reaching the large-scale capacities required for fully practical quantum computers and networks. 
As a result, there has been considerable research into emulating quantum algorithms on classical computing resources to bridge the gap until fully operational QC systems emerge. 
QC emulators play a critical role in enabling rapid experimentation and prototyping of quantum circuits, traditionally relying on large, resource-intensive platforms~\cite{Arrazola:Nature21}. In contrast, recent QC emulators increasingly leverage FPGAs, providing more efficient, scalable, and cost-effective solutions for hardware-accelerated quantum algorithm emulation~\cite{Shang:npjQI23, ElAraby:IEEE23,Choi:AIMS24}. 
The inherent cost-efficiency, high performance, and reconfigurability of FPGAs make them an appealing choice for QC emulation.

%


Nevertheless, most FPGA-based QC emulators to date are designed as QC accelerators integrated into high-performance computing systems \cite{Pilch:Springer19, Li:IEEE21, Shang:npjQI23, ElAraby:IEEE23} . 
While such integration is valuable, it falls short of supporting the eventual deployment of QC applications in mobile and edge device environments. 
Mobile and edge devices typically operate under stringent resource constraints and require applications that can closely interact with diverse surrounding environments and situations. 
Consequently, a QC emulator that operates within these constraints on embedded devices, rather than relying on server-level computing power, is essential.

In response to this need, we aim to develop a QC emulator tailored for embedded devices, designed to maximize efficiency within limited resources. 
Our research focuses on designing an optimized QC accelerator and implementing a complete standalone system on an FPGA. 
We have selected the Quantum Approximate Optimization Algorithm (\textit{QAOA}) as our target algorithm, given its high potential for applications in mobile and edge device environments. 
QAOA is an effective QC method for finding approximate solutions to combinatorial optimization problems~\cite{Streif:PRA21, Borle:SciPost21, Moussa:Springer22, Awasthi:Springer23} and is particularly relevant for the MaxCut problem in graph partitioning~\cite{Blekos:Elsevier24, Zhao:Elsevier24, Esposito:IEEE24}. 
Graph partitioning optimization is essential in numerous embedded system applications, especially in resource-constrained environments where efficient partitioning algorithms are critical~\cite{Jungum:Elsevier16, Faraj:ACM22, Chhabra:arXiv24}. 
Our study on an FPGA standalone QAOA emulator is expected to serve as a pivotal benchmark, providing critical insights and establishing a foundational framework for the practical deployment and advancement of QC within embedded systems.

The primary challenge in developing a QAOA emulator lies in creating an accelerator that supports the maximum number of qubits within constrained resources. 
Increasing the number of qubits in QC enhances the emulator's computational power and capacity, emphasizing the need for optimized design techniques. 
Like other QC algorithms, QAOA represents quantum states and operations as vectors and matrices, which grow exponentially with the number of qubits $n$. This exponential growth demands substantial resources for storage and processing, posing significant challenges for QC emulation on resource-constrained platforms. 
To address these challenges, we adopt a hardware-software co-design approach. 
First, we optimize the computational expressions of QAOA for the Weighted-MaxCut problem, tailoring them for hardware implementation and reducing the emulator's time complexity from $O(N^2)$ to $O(N)$ ($N=2^n$). 
We then deploy these optimized expressions in a pipelined structure, minimizing hardware resource requirements by reducing the number of parallelized multiplications, which are particularly resource-intensive in hardware computations.


Finally, the proposed accelerator was implemented as an IP core and integrated into an edge-device-oriented RISC-V processor architecture, completing the standalone QC emulator. 
Performance evaluations reveal that the energy efficiency of the emulator, enhanced by the QAOA accelerator, increases exponentially with the number of qubits, achieving up to a 1.5$\times$ reduction for 2-qubit QAOA and reaching a maximum of 854$\times$ reduction for 9-qubit QAOA.
Resource utilization analyses conducted on FPGA prototypes demonstrate that, while previous approaches support accelerators for up to 3-qubits on mid-tier FPGAs (e.g., Kintex-7) and 2-qubits on entry-level boards (e.g., Artix-7), our resource-optimized design enables the implementation of QC accelerators supporting up to 9-qubits and 8-qubits, respectively. 
These results highlight substantial improvements in both resource and energy efficiency, establishing the proposed emulator as a practical and scalable QC solution for solving complex Weighted-MaxCut problems on resource-constrained embedded devices.

\begin{figure}
	\centering
	\includegraphics[width=0.49\textwidth]{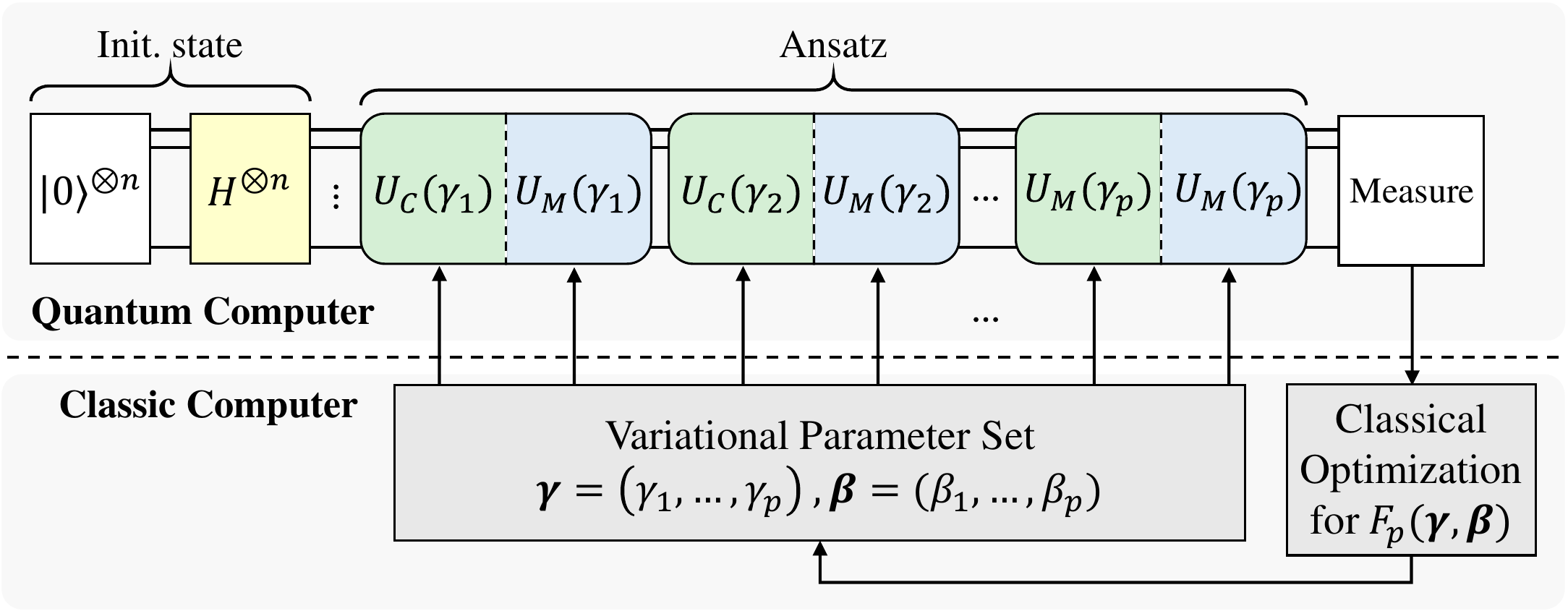}
	\caption{Overall schematic of the QAOA process.}
	\label{fig:qaoa}
\end{figure}

\section{QAOA for Weighted-MaxCut: Preliminaries}

\subsection{Quantum Approximate Optimization Algorithm} 
%
%

As shown in \refFigure{fig:qaoa}, the operations in QAOA consist of a pair of \textit{ansatz} operators: the cost ansatz operator $U_{C}(\gamma_{k})$ and the mixer ansatz operator $U_{M}(\beta_{k})$. 
In a QAOA circuit of depth $p$, the two ansatz operators in the $k^{\text{th}}$ layer $(k \leq p)$ are adjusted by the parameters $\gamma_{k}$ and $\beta_{k}$ of the QAOA circuit, as represented by the following equations~\cite{Farhi:arXiv14}:
\begin{equation}\label{eq:ansatz_op_pair}
    U_{C}(\gamma{}_{k}) = e^{-i \gamma{}_{k} \hat{H}_{C}}, \quad{} U_{M}(\beta{}_{k}) = e^{-i \beta{}_{k} \hat{H}_{M}},
\end{equation}
\normalsize
where $\hat{H}_{C}$ denotes the cost Hamiltonian representing the cost function of the optimization problem, and $\hat{H}_{M}$ is the mixer Hamiltonian that induces superposition in quantum states, thereby expanding the search space.

The QAOA circuit sequentially applies the ansatz operator pairs defined in (\ref{eq:ansatz_op_pair}) for each $k$ from $1$ to $p$, on an initial quantum state $\vert s \rangle = H^{\otimes n} \vert 0 \rangle^{\otimes n}$. 
A total of $2p$ variational parameters, $\boldsymbol{\gamma} = (\gamma_{1}, \ldots, \gamma_{p})$ and $\boldsymbol{\beta} = (\beta_{1}, \ldots, \beta_{p})$, are used to tune the operations of each ansatz.

After completing the $p$ layers of ansatz operations, the resulting ansatz state $\vert \psi_{p}(\boldsymbol{\gamma}, \boldsymbol{\beta}) \rangle$ is expressed as follows:
\begin{equation}
    \vert{}\psi{}_{p}(\boldsymbol{\gamma{}}, \boldsymbol{\beta{}})\rangle{} = e^{-i \beta{}_{p} \hat{H}_{M}} e^{-i \gamma{}_{p} \hat{H}_{C}} \ldots e^{-i \beta{}_{1} \hat{H}_{M}} e^{-i \gamma{}_{1} \hat{H}_{C}} \vert{}s\rangle{}.
\end{equation}
\normalsize
To ensure that the ansatz state obtained through this operation yields a solution close to the optimal, it is essential to find the optimal set of parameters $(\boldsymbol{\gamma}, \boldsymbol{\beta})$. 
For this purpose, on the classical computing side, the ansatz state output by the quantum computer is repeatedly measured, and the expectation value $F_{p}(\boldsymbol{\gamma}, \boldsymbol{\beta})$ of the cost Hamiltonian $\hat{H}_{C}$ is computed. 
This expectation value, corresponding to the cost function of the given problem, can be expressed in both quantum mechanical and probabilistic representations as follows:
\begin{equation}\label{eq:expect_prob}
    F_{p}(\boldsymbol{\gamma}, \boldsymbol{\beta}) = \langle \psi_{p}(\boldsymbol{\gamma}, \boldsymbol{\beta}) \vert \hat{H}_{C} \vert \psi_{p}(\boldsymbol{\gamma}, \boldsymbol{\beta}) \rangle = \sum\nolimits_{x} p(x) C(x)
\end{equation}
where $p(x)$ denotes the probability of observing solution $x$ in measurement samples of the ansatz state. 
On the classical computer, an optimization algorithm is employed to maximize the expectation value obtained from (\ref{eq:expect_prob}), thereby finding the parameter set $(\boldsymbol{\gamma}, \boldsymbol{\beta})$ that maximizes this expectation value. On the quantum computer, the QAOA algorithm is re-executed based on the updated parameter set, iteratively optimizing the ansatz state to gradually converge towards the optimal solution.

\subsection{Weighted-MaxCut Problem}

\begin{figure}
	\centering
	\includegraphics[width=0.45\textwidth]{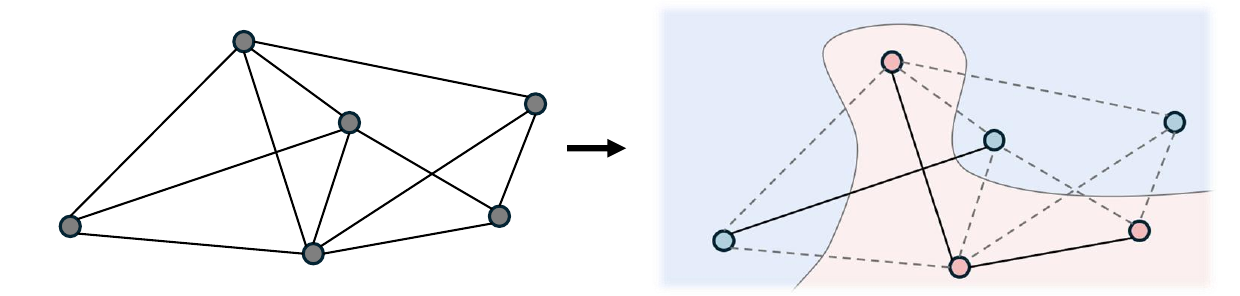}
	\caption{Example of a Weighted-MaxCut problem with 6 vertices and 11 edges (weights omitted for simplicity).}
	\label{fig:maxcut}
\end{figure}

The Weighted-MaxCut problem involves partitioning a graph with given vertices, edges, and edge weights in a way that maximizes the total weight of the edges crossing the partition.
\refFigure{fig:maxcut} presents a graphical representation of a Weighted-MaxCut problem, where the partition shown on the right maximizes the number of cut edges or their total weight, as defined by the problem on the left.
More specifically, each vertex $x_i$ in the $i$-th position belongs to one of the two subsets formed by the partition and can be labeled as 0 or 1 accordingly.
Thus, for a vertex set of size $\vert V \vert$, any partition can be represented as a bit string $\mathbf{x} = x_{1} x_{2} \ldots x_{\vert V \vert}$.
By defining a cost function $C(\mathbf{x})$ that takes $\mathbf{x}$ as input and represents the total weight of the cut, the Weighted-MaxCut problem can be formulated as the task of finding the bit string $\mathbf{x}$ that maximizes $C(\mathbf{x})$.
The cost function $C(\mathbf{x})$ for the Weighted-MaxCut problem is given by:
\begin{equation}
    C(\mathbf{x}) = \sum_{i, j = 1}^{\vert{}V\vert{}} \omega{}_{ij} x{}_{i} (1 - x{}_{j}).
\end{equation}
If the $i$-th and $j$-th vertices belong to the same subset, then $(x_{i}, x_{j}) = (0,0)$ or $(1,1)$, resulting in $x_{i} (1 - x_{j}) = 0$. 
Consequently, the edge connecting these two vertices does not affect the value of the cost function. 
In contrast, if the two vertices are in different subsets ($(x_{i}, x_{j}) = (0,1)$ or $(1,0)$), then $x_{i} (1 - x_{j})$ is nonzero, and this edge contributes to the cost function.

To solve the Weighted-MaxCut problem using QAOA, the cost Hamiltonian and mixer Hamiltonian must be defined. 
For the set of edges $E$ contributing to the cost function, the two Hamiltonians for the Weighted-MaxCut problem are defined as follows:

\begin{equation}\label{eq:cost_hamiltonian}
    \hat{H}_{C} = \frac{1}{2} \sum_{(i,j) \in E} \omega{}_{ij} (I - Z_{i} Z_{j}),~~\hat{H}_{M} = \sum_{j \in V} X_{j},
\end{equation}
where $Z_{i}$ is the Pauli-Z gate operation on the $i$-th qubit, and the $l$-th diagonal element $Z_{i}Z_{j}(l,l)$ of the term $Z_{i} Z_{j}$ can be expressed as:
\begin{equation}\label{eq:ZiZj_diag}
    Z_{i}Z_{j}(l,l) = \begin{cases}
    1 & \text{if the $l$-th bits of $i$ and $j$ are different} \\
    -1 & \text{if the $l$-th bits of $i$ and $j$ are the same}
    \end{cases}
\end{equation}

To utilize the Hamiltonians constructed for the Weighted-MaxCut problem within the QAOA circuit, they must be transformed into unitary quantum gates. For a circuit depth of $p$, the cost unitary and mixer unitary operations in the $k$-th layer $(k \leq p)$ for the Weighted-MaxCut problem can be derived as follows:
\begin{equation}\label{eq:cost_unitary}
    U_{C}(\gamma_{k}) = e^{-i \gamma_{k} \hat{H}_{C}} = \prod_{i=1,j<i}^{n} R_{Z_{i} Z_{j}}(-2 \omega_{ij} \gamma_{k}).
\end{equation}
\begin{equation}\label{eq:mix_unitary}
    U_{M}(\beta_{k}) = e^{-i \beta_{k} \hat{H}_{M}} = \prod_{i=1}^{n} R_{X_{i}}(2 \beta_{k}).
\end{equation}

\section{Optimization Strategies for the QAOA Emulator}

To emulate QAOA, it is necessary to repeatedly multiply the unitary matrices representing the ansatz operations, $U_{C}(\gamma_{k})$ and $U_{M}(\beta_{k})$, with the vector representing the initial state. 
However, for emulating $n$-qubit QAOA, the multiplication of the $N \times N$ matrix and $N \times 1$ vector (where $N=2^n$, the state space dimension) requires registers and computing power that scale with $O(N^2)$ ($=O(2^{2n})$) as $n$ increases, leading to an exponential growth in resource demands. 
Therefore, to efficiently emulate QAOA within the limited hardware resources of an SoC, it is essential to modify the matrix representation of QAOA in a hardware-friendly manner and accelerate the modified matrix operations by leveraging hardware characteristics. 
In this study, we decompose the matrix representation of the mixer ansatz operation $U_{M}(\beta_{k})$ for the Weighted-MaxCut problem, structuring the entire ansatz operation into hardware-optimized elemental operations. 
Furthermore, a pipelined structure is employed for these elemental operations to reduce the hardware resource demands associated with parallelizing multiplication operations.


\subsection{Decomposing the Mixer Unitary}\label{sec:decomp_mix_unitary}

The matrix representation of the Pauli-X rotation gate $R_{X}(2\beta_{k})$, which defines the mixer ansatz in (\ref{eq:mix_unitary}), is given by

\begin{equation}
\begin{aligned}
    R_{X}(2\beta_{k}) &= e^{-i\beta_{k} X} = \cos(\beta_{k}) I - i \sin(\beta_{k}) X \\
    &= \begin{bmatrix}
    \cos(\beta_{k}) & -i\sin(\beta_{k}) \\
    -i\sin(\beta_{k}) & \cos(\beta_{k})
    \end{bmatrix}.
\end{aligned}
\end{equation}
In this equation, both the identity matrix $I$ and the Pauli matrix $X$ have the vectors $\mathbf{v_{1}} = \begin{bmatrix} 1 & 1 \end{bmatrix}^{T}$ and $\mathbf{v_{2}} = \begin{bmatrix} 1 & -1 \end{bmatrix}^{T}$ as eigenvectors. 
Therefore, $R_{X}(2\beta_{k})$ also has $\mathbf{v_{1}}$ and $\mathbf{v_{2}}$ as eigenvectors, with eigenvalues determined as follows:
\begin{equation}\notag{}
    R_{X}(2\beta{}_{k})  \mathbf{v_{1}} = \begin{bmatrix}
    \cos(\beta{}_{k}) & -i\sin(\beta{}_{k}) \\
    -i\sin(\beta{}_{k}) & \cos(\beta{}_{k})
    \end{bmatrix}
     \begin{bmatrix} 1 \\ 1 \end{bmatrix} \qquad{} \qquad{}\qquad{}
\end{equation}
\begin{equation}
    \qquad{}= (\cos(\beta{}_{k}) - i \sin(\beta{}_{k})) \begin{bmatrix} 1 \\ 1 \end{bmatrix}
    = e^{-i\beta{}_{k}}  \mathbf{v_{1}} = \lambda{}_{1}  \mathbf{v_{1}},
\end{equation}
\begin{equation}\notag{}
    R_{X}(2\beta{}_{k})  \mathbf{v_{2}} = \begin{bmatrix}
    \cos(\beta{}_{k}) & -i\sin(\beta{}_{k}) \\
    -i\sin(\beta{}_{k}) & \cos(\beta{}_{k})
    \end{bmatrix}
    \begin{bmatrix} 1 \\ -1 \end{bmatrix} \qquad{}\qquad{}\qquad{}
\end{equation}
\begin{equation}
   \qquad{} = (\cos(\beta{}_{k}) + i \sin(\beta{}_{k})) \begin{bmatrix} 1 \\ -1 \end{bmatrix}
    = e^{i\beta{}_{k}} \mathbf{v_{2}} = \lambda{}_{2} \mathbf{v_{2}},
\end{equation}
where $\lambda_{1} = e^{-i\beta_{k}}$ and $\lambda_{2} = e^{i\beta_{k}}$. 
Thus, the eigenvalue decomposition of $R_{X}(2\beta_{k})$ is as follows:
\begin{equation}\label{eq:hadamard_rx_diag}
    R_{X}(2\beta{}_{k}) = \begin{bmatrix} 1 & 1 \\ 1 & -1 \end{bmatrix}
    \begin{bmatrix} e^{-i\beta{}_{k}} & 0 \\ 0 & e^{i\beta{}_{k}} \end{bmatrix}
    \begin{bmatrix} 1 & 1 \\ 1 & -1 \end{bmatrix}^{-1}
    = H \Lambda{}_{M} H,
\end{equation}
where $H$ is the Hadamard matrix, and $\Lambda_{M}$ is the diagonal matrix containing the eigenvalues of $R_{X}(2\beta_{k})$ as its diagonal elements.

It can be seen that $U_{M}(\beta_{k})$ in (\ref{eq:mix_unitary}) represents an operation that applies the $R_{X}(2\beta_{k})$ operation simultaneously to $n$ qubits. 
Therefore, $U_{M}(\beta_{k})$ can be expressed as the tensor product of the $R_{X}(2\beta_{k})$ matrices as follows:
\begin{equation}\label{eq:mix_eigen_diag}
    U_{M}(\beta{}_{k}) = R_{X}(2\beta{}_{k})^{\otimes n} = (H \Lambda{}_{M} H)^{\otimes n} = H^{\otimes n} D_{M} H^{\otimes n},
\end{equation}
where $D_{M}$ is the diagonal matrix representing the mixer ansatz and is given by $\Lambda_{M}^{\otimes n}$.

In addition, in (\ref{eq:cost_hamiltonian}), the cost Hamiltonian $\hat{H}_{C}$ for the Weighted-MaxCut problem is a diagonal matrix, as it is the partial sum of diagonal matrices $I$ and $Z_{i} Z_{j}$. 
Consequently, the matrix exponential $U_{C}(\gamma_{k})$ in (\ref{eq:cost_unitary}), which has $\hat{H}_{C}$ as its exponent, is also a diagonal matrix. 
Defining the diagonal matrix representing the cost ansatz as $D_{C}$, i.e., $D_{C} = U_{C}(\gamma_{k})$, the ansatz operation performed in the $k$-th layer can be expressed as follows:
\begin{equation}\label{eq:unit_ansatz_op}
    U_{M}(\beta{}_{k}) U_{C}(\gamma{}_{k}) = (H^{\otimes n} D_{M})(H^{\otimes n} D_{C}).
\end{equation}
In this equation, the ansatz operation for a single layer consists of an elemental ansatz operation ($U_{A} = H^{\otimes n} D$), which is the product of a diagonal matrix and the Hadamard matrix, applied separately for the cost and mixer operations. 
Therefore, by computing the values of the diagonal matrices $D_{M}$ and $D_{C}$ for all layers, the ansatz state can be obtained by iteratively applying these operations to the initial quantum state $\vert s \rangle$. 
Importantly, note that the computational cost for the diagonal matrix operations in this decomposed form is $O(N)$, which is significantly lower than the $O(N^{2})$ required for the original matrix-vector multiplication.

Meanwhile, in (\ref{eq:unit_ansatz_op}), the Hadamard matrix $H^{\otimes n}$ has all elements with an absolute value of $1/\sqrt{2^{n}}$. 
Therefore, by scaling the original Hadamard matrix to construct a modified Hadamard matrix $H_{1}$, in which all elements have an absolute value of 1, the multiplication operations required for the Hadamard transform in the elemental ansatz operation can be reduced. 
The relationship between the two matrices is given as follows:
\begin{equation}
    H_{1} = \frac{1}{\sqrt{2^{n}}} H^{\otimes n}, \quad \text{or} \quad H^{\otimes n} = \sqrt{2^{n}} H_{1}.
\end{equation}
\normalsize
Substituting this into (\ref{eq:unit_ansatz_op}) and simplifying yields the following:
\begin{equation}\label{eq:unit_ansatz_op_scaled}
    U_{M}(\beta{}_{k}) U_{C}(\gamma{}_{k}) = \frac{1}{2^{n}} (H_{1} D_{M})(H_{1} D_{C}).
\end{equation}

The modified Hadamard transform can be implemented resource-efficiently by using only parallel addition/subtraction operations on vector components, instead of multiplication and accumulation operations. 
The scaling factor in powers of $2$ in (\ref{eq:unit_ansatz_op_scaled}) can be applied in hardware with simple bit-shift operations.

\subsection{Constructing the Cost and Mixer Ansatz Diagonal Matrices}
As derived in the previous section, the cost Hamiltonian matrix for the MaxCut problem is a diagonal matrix. 
Thus, by leveraging this diagonal property, the diagonal elements of the cost ansatz matrix $D_{C}$ can be derived from the cost Hamiltonian. 
First, using the definition of the matrix exponential and (\ref{eq:cost_unitary}), $D_{C}$ can be expressed as follows:
\begin{equation}\label{eq:diag_cost}
    D_{C} = U_{C}(\gamma{}_{k}) = e^{-i \gamma{}_{k} \hat{H}_{C}} = \sum_{k=0}^{\infty} \frac{(-i \gamma{}_{k})^{m}}{m!} \hat{H}_{C}^{m}
\end{equation}
Since $\hat{H}_{C}$ is a diagonal matrix, the $l$-th diagonal element of its power $\hat{H}_{C}^{m}$, denoted as $\hat{H}_{C}^{m}(l,l)$ for $l = 1, 2, \ldots, n$, can be obtained as follows:
\begin{equation}
    \hat{H}_{C}^{m}(l,l) = (\hat{H}_{C}(l,l))^{m}
\end{equation}
Thus, by generalizing the $l$-th diagonal element $D_{C}(l,l)$, $D_{C}$ can be constructed for a given cost Hamiltonian as follows:
\begin{equation}
  D_{C}(l,l) = \sum_{m=0}^{\infty} \frac{(-i \gamma{}_{k} \hat{H}_{C}(l,l))^{m}}{m!} = e^{-i \gamma{}_{k} \hat{H}_{C}(l,l)}
\end{equation}

Next, we derive a general expression for the $l$-th diagonal element $D_{M}(l,l)$ of the diagonal matrix $D_{M}$ representing the mixer ansatz, for $l = 1, 2, \ldots, n$. 
Using (\ref{eq:hadamard_rx_diag}) and (\ref{eq:mix_eigen_diag}), we know that $D_{M}$ can be represented as the tensor product of the $2 \times 2$ diagonal matrix $\Lambda_{M}$. 
Therefore, each diagonal element of $D_{M}$ is expressed as the successive product of the two diagonal elements of $\Lambda_{M}$, namely $\lambda_{1} = e^{-i\beta_{k}}$ and $\lambda_{2} = e^{i\beta_{k}}$. 
To compute the $l$-th diagonal element of the matrix $D_{M}$, each of the two diagonal elements of $\Lambda_{M}$ is multiplied $n - HW(l-1)$ times and $HW(l-1)$ times, respectively, where $HW(l-1)$ denotes the Hamming weight of $l-1$, which is the number of 1’s in the binary representation of $l-1$. 
Based on this, $D_{M}(l,l)$ can be expressed as follows:
\begin{equation}\label{eq:diag_mix}
    D_{M}(l, l) = \lambda{}_{1}^{n-HW(l-1)} \times{} \lambda{}_{2}^{HW(l-1)} = e^{iu(n,l)\beta{}_{k}},
\end{equation}
where $u(n,l) = 2 \cdot HW(l-1) - n$. 

\subsection{Pipelining Elemental Ansatz Operations}\label{sec:pipeline_unit_ansatz_op}

\begin{figure}
	\centering
	\includegraphics[width=0.42\textwidth]{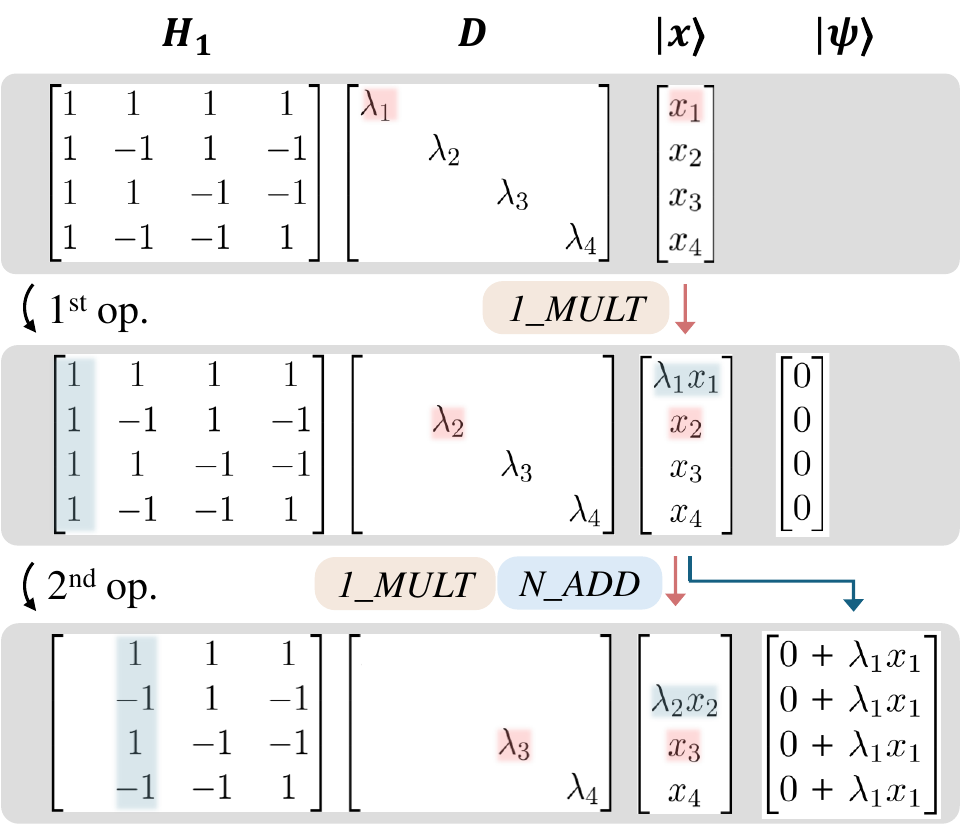}
	\caption{Operations in the pipeline structure for a single elemental ansatz on a 2-qubit system.}
	\label{fig:pipeline}
\end{figure}

Through the proposed matrix decomposition and scaling methods, we structured the computations required for the QAOA algorithm emulation for the MaxCut problem into hardware-friendly elemental ansatz operations, and derived general expressions for the diagonal elements of the diagonal matrices $D_{C}$ and $D_{M}$ required for these operations. 
However, to perform the multiplication of the diagonal matrix and vector in parallel within these elemental operations, an exponentially increasing number of multiplications would still be required. 
As is well known, multiplication operations demand significantly more hardware resources and computational power compared to addition operations, so parallelizing these exponentially increasing multiplications places a substantial burden on hardware.

To address this issue, we process these diagonal matrix operations in a pipeline in conjunction with the Hadamard matrix operations, thereby minimizing the parallelization of multiplication operations and implementing a hardware-efficient emulation algorithm. 
\refFigure{fig:pipeline} illustrates the proposed pipeline design.
In the first operation (denoted as \textit{1\textsuperscript{st}.op} in the figure), instead of performing parallel multiplications for all diagonal elements of the matrix $D$ with the quantum state vector $\vert x \rangle$, we perform a single multiplication per clock cycle, focusing on one element at a time. 
This is represented by \textit{\textit{1\_MULT}} in the figure.

The result of the first multiplication operation, $\lambda_{1} x_{1}$, corresponds to the first column element of the modified Hadamard matrix $H_{1}$. 
In the subsequent clock cycle, the \textit{2\textsuperscript{nd}.op} operation adds or subtracts this result to the output vector $\vert \psi \rangle$ based on the column element's sign (either $1$ or $-1$), which is denoted as \textit{N add.} in the figure.
Simultaneously, in \textit{2\textsuperscript{nd}.op} of the same clock cycle, the next multiplication (\textit{1\_MULT}) is performed between the second element of $D$ and $\vert x \rangle$. 
In this pipeline design, each clock cycle within a single elemental ansatz operation executes one multiplication and $N$ additions in parallel. 
This sequence of parallel operations iterates across all $N$ elements of the diagonal matrix and the quantum state vector over a total of $N+1$ clock cycles to compute the ansatz state.

This pipeline architecture significantly reduces the computation time of the QAOA emulator’s matrix operations, decreasing the complexity from $O(N^{2})$ to $O(N)$ through efficient hardware parallelization. 
Additionally, by minimizing the need for parallel multiplications—which demand far more hardware resources than additions—our design mitigates the exponential increase in hardware requirements that typically accompanies larger qubit counts. 
By controlling the number of multiplication operations, we balance the reduction in execution time with manageable hardware resource usage. 
As a result, this optimized pipeline design effectively limits the growth in hardware resource demand, enabling scalable and resource-efficient QAOA emulation for the Weighted-MaxCut problem.

\section{Design and Implementation of the QAOA Emulator}

\subsection{QAOA Emulator Architecture and Design Automation}

\begin{figure}
	\centering
	\includegraphics[width=0.5\textwidth]{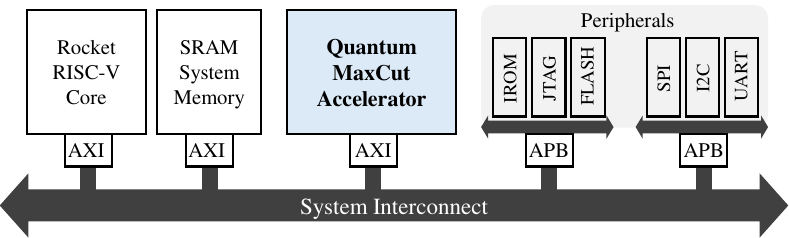}
	\caption{Architecture diagram of the developed emulator with the Quantum MaxCut Accelerator (QMA).}
	\label{fig:architecture}
\end{figure}

To implement the QAOA emulator for the Weighted-MaxCut problem on standalone FPGAs, we designed a RISC-V processor platform based on the architecture shown in \refFigure{fig:architecture}. 
This platform leverages the RISC-V Rocket core~\cite{Berkeley:Rocket}, which includes floating-point arithmetic units, making it well-suited for QAOA emulation in embedded systems. 
Key components of the architecture include: a 512KB SRAM as the main memory, Flash memory for non-volatile storage, 
a lightweight Network-on-Chip ($\mu NoC$), a control module for boot and reset functions, and standard I/O interface modules (UART, SPI, and I$^{2}$C).
For the RTL design, we utilized the RISC-V eXpress (\textit{RVX})\cite{Han:IEEE21}, an EDA tool widely adopted for lightweight RISC-V processor development~\cite{Park:TCASI24, Lee:IoTJ24, Han:ISLPED24, Choi:JETECH2024}. 
The system was configured to operate at a 100MHz clock frequency. 
Central to this design is the Quantum MaxCut Accelerator (\textit{QMA}), a dedicated hardware module that integrates our proposed resource optimization techniques for efficient QAOA processing on the Weighted-MaxCut problem.
The QMA employs fixed-point arithmetic, which is well-suited for hardware-level operations~\cite{Zhang:FPGA15}, and is developed as an IP core with AXI interface support. 
This design enables seamless integration into any processor platform. 
As shown in \refFigure{fig:architecture}, we embedded this IP core into the baseline RISC-V processor, resulting in an emulator capable of efficiently executing QAOA operations with minimal additional resource usage.

To automate the emulator’s RTL code generation for varying qubit counts, we extended RVX to develop QC Emulator eXpress (\textit{QEX}). 
QEX accepts user-defined parameters, 
\textit{NUM\_QUBIT}, representing the target number of qubits. 
Given \textit{NUM\_QUBIT}, QEX determines the required register blocks for storing coefficients of basis states and elements of diagonal matrices, defining this count as a local parameter, \textit{NUM\_STATE}. 
QEX then configures the registers and computation units of the MaxCut Accelerator accordingly, generating the complete RTL code of the SoC platform depicted in \refFigure{fig:architecture}.

We further developed a C-language API to interface with the QMA, providing four primary commands: 
i) \texttt{set\_parameter}, 
ii) \texttt{set\_cost\_hamiltonian}, 
iii) \texttt{activate\_maxcut}, and 
iv) \texttt{get\_expectation}. 
The emulator operates by storing application instructions, implemented with these API commands, in the SRAM for execution by the Rocket core. Upon completing the emulation, the MaxCut Accelerator returns the ansatz state’s expectation value to the Rocket core, enabling the user to optimize variational parameters based on the outcome.

%
%

\begin{figure*}
	\centering
	\includegraphics[width=1\textwidth]{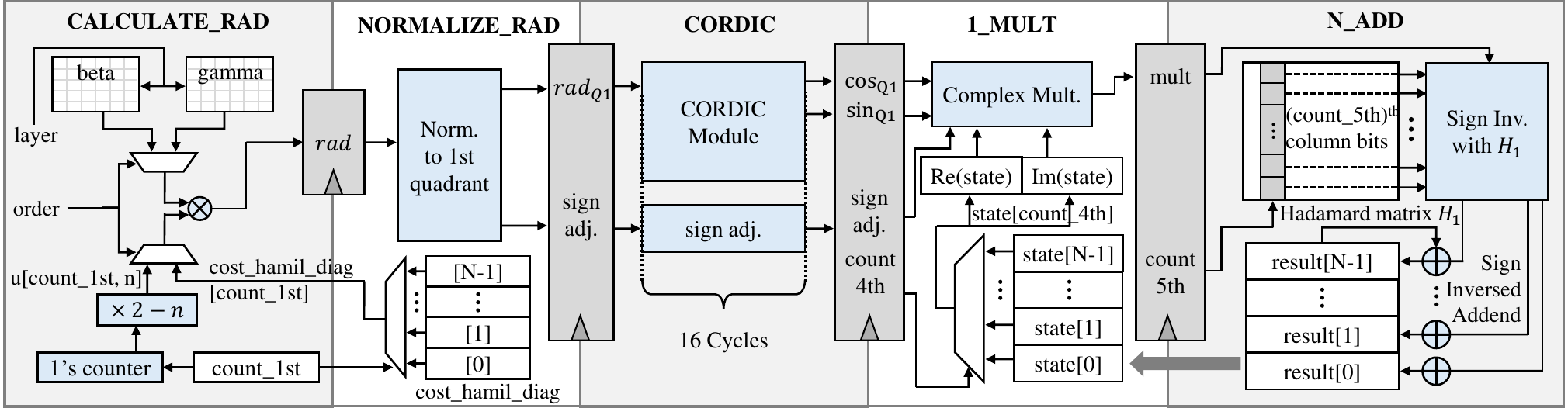}
	\caption{Register-level schematic of the pipeline stages in the developed QMA.}
	\label{fig:reg_pipeline}
\end{figure*}

\subsection{Setup for Emulating QAOA}
Before running QAOA, it is necessary to define the cost Hamiltonian for the Weighted-MaxCut problem. 
To do this, the two vertices $i$ and $j$ connected by each edge in the graph, along with the fixed-point representation of the edge weight $weight$, are input into the QMA using the \texttt{set\_cost\_hamiltonian} command. 
The Hamiltonian is then constructed based on (\ref{eq:cost_hamiltonian}) and (\ref{eq:ZiZj_diag}), representing it as a partial sum of $Z_{i}Z_{j}$ matrices. 
Its diagonal elements, \texttt{cost\_hamil\_diag}, are calculated according to the following algorithm:

\fontsize{7}{7}\selectfont{}
\begin{verbatim}
 for (k = 0; k < NUM_STATE; k++) {
   bit_i = (k >> (i-1)) & 1;   bit_j = (k >> (j-1)) & 1;
   if (bit_i != bit_j) cost_hamil_diag[k] += 2.0 * weight;}
\end{verbatim}
\normalsize

Additionally, the \texttt{set\_parameter} command configures the QAOA layer number (\textit{layer}) and sets the cost and mixer ansatz parameters in fixed-point format. 
These parameters are stored in the \textit{gamma[layer]} and \textit{beta[layer]} registers, respectively. 
The \textit{NUM\_QUBIT} parameter adjusts the number of qubits available for QAOA, defining the size of the Weighted-MaxCut problem that the emulator can handle when synthesized on the FPGA. 
This setup ensures that the QMA scales efficiently with the problem size specified by the user.

%
%
%

\subsection{Register-Level Pipeline}

In the pipeline structure described in Section~\ref{sec:pipeline_unit_ansatz_op}, the diagonal elements of $D_{c}$ and $D_{m}$ are required. 
These elements are computed as the polar form exponents (i.e., angle values) of each Hamiltonian, determined based on the parameters using (\ref{eq:diag_cost}) and (\ref{eq:diag_mix}). 
To perform computations on these elements in hardware, the angle values must be converted to their Cartesian form through trigonometric function evaluation. 
In this study, we extended the pipeline described in Section~\ref{sec:pipeline_unit_ansatz_op} by concatenating a CORDIC module~\cite{Volder:IEEE59} as an additional stage for trigonometric calculations, thereby constructing a complete pipeline for computing elemental ansatz operations based on the specified cost Hamiltonian and parameters.
\refFigure{fig:reg_pipeline} illustrates the register-level schematic for each stage of the pipeline.

This pipeline is executed $2p$ times across $p$ layers, where input values change based on the layer index and the sequence of $D_{c}$ and $D_{m}$ to ensure correct computation of each elemental ansatz operation. To control this, we defined a register \textit{order} to indicate whether each ansatz operation corresponds to the cost or mixer unitary, as well as a register \textit{layer} to specify the current layer index.

%

\subsubsection{\textbf{CALCULATE\_RAD} stage}\noindent
This stage corresponds to the first stage depicted in \refFigure{fig:reg_pipeline}.
In this stage, the radian angle values of each diagonal element in $D_{c}$ and $D_{m}$ are computed based on (\ref{eq:diag_cost}) and (\ref{eq:diag_mix}).
When the \texttt{activate\_maxcut} command is executed, the pipeline stages operate over \textit{NUM\_STATE} clock cycles. 
The counter register \textit{count\_1st} ensures that each input corresponds to the (\textit{count\_1st})-th diagonal element of the Hamiltonian.

During this stage, the inputs are defined as the product of \textit{cost\_hamil\_diag[count\_1st]} and \textit{gamma[layer]} for $D_{c}$, or, for $D_{m}$, the product of $\textit{u[count\_1st, NUM\_QUBIT]}$ (cf. (\ref{eq:diag_mix})) and \textit{beta[layer]}. These inputs are selected based on the values of \textit{order} and \textit{layer}. The resulting angle values are stored in the \textit{rad} register and forwarded through the pipeline. The assignment of the \textit{rad} follows the algorithm:

\fontsize{7}{7}\selectfont
\begin{verbatim}
if(!order) {rad = cost_hamil_diag[count_1st] * gamma[layer];}
else {rad = u(count_1st, NUM_QUBIT) * beta[layer];}
\end{verbatim}
\normalsize

\subsubsection{\textbf{NORMALIZE\_RAD} stage}\noindent
To ensure accurate trigonometric calculations in the CORDIC module, the angle values computed in the \textit{CALCULATE\_RAD} stage are normalized to the first quadrant as illustrated in \refFigure{fig:reg_pipeline}. 
First, the given angle value is reduced modulo $2\pi$, and then the normalized angle is calculated based on the quadrant in which the angle lies. 
This normalized angle is stored in the $rad_{Q1}$ register. Additionally, the signs of the trigonometric values, which may change due to quadrant normalization, are stored in the \textit{sign adjustment} registers based on the original quadrant of \textit{rad}. 
The angle values and trigonometric signs are determined according to the following algorithm:

\fontsize{7}{7}\selectfont{}
\begin{verbatim}
 neg_cos = 0; neg_sin = 0; norm_rad = rad % (2 * PI);
 if(en_pipeline) {
   if(norm_rad < 0.5 * PI) norm_rad_first_quad = norm_rad;
   else if(norm_rad < PI) 
     norm_rad_first_quad = PI - norm_rad; neg_cos = 1;
   else if(norm_rad < 1.5 * PI) {
     norm_rad_first_quad = norm_rad - PI;
     neg_sin = 1; neg_cos = 1;}
   else {
     norm_rad_first_quad = 2 * PI - norm_rad; neg_sin = 1;}}  
\end{verbatim}

\normalsize

%

\normalsize

\subsubsection{\textbf{CORDIC} stage} \noindent

This stage employs a CORDIC module to compute both cosine and sine values of $rad_{Q1}$.
The CORDIC module is implemented as a 16-stage pipeline, allowing the input angle $rad_{Q1}$ to propagate through the pipeline over 16 clock cycles.
After these cycles, the CORDIC module outputs the trigonometric values ${cos}_{Q1}$ and ${sin}_{Q1}$, along with a corresponding valid signal indicating completion.
To align the sign adjustments \delete{(\textit{neg_sin} and \textit{neg_cos})} with the trigonometric outputs generated in the same clock cycle, these sign values are pipelined in parallel with the CORDIC module across all 16 stages.



\subsubsection{\textbf{1\_MULT} stage}\noindent
In this stage, the diagonal matrix element obtained from the \textit{CORDIC} stage is multiplied by the corresponding component of the state vector, performing the multiplication operation represented as \textit{1\_MULT} in \refFigure{fig:pipeline}. 
As shown in the \textit{1\_MULT} stage of \refFigure{fig:reg_pipeline}, a register, \textit{count\_4th}, is defined in this stage to count the number of operations. 
This counter selects the \textit{state[count\_4th]} from the \textit{state} vector, which represents the quantum states, and proceeds with the multiplication together with the diagonal element computed in the previous stage. 
Since both the diagonal matrix element and the \textit{state} vector component are complex numbers, the real and imaginary parts of the operation are calculated separately, as shown in the following algorithm:


\fontsize{7}{7}\selectfont{}
\begin{verbatim}
    Re(mult) = Re(state[count_4th]) * cos_theta
        - Im(state[count_4th]) * sin_theta;
    Im(mult) = Re(state[count_4th]) * sin_theta
        + Im(state[count_4th]) * cos_theta;
\end{verbatim}
\normalsize
%
%

As a result, the real and imaginary parts of the complex product of the two complex numbers are output, and these values are stored in the \textit{mult} register to be passed to the next stage.

\subsubsection{\textbf{N\_ADD} stage}\noindent
As shown in \refFigure{fig:reg_pipeline}, this stage takes the complex product obtained in the \textit{1\_MULT} stage and multiplies it by each element in the corresponding column of the Hadamard matrix ($H_1$), performing the accumulation operation on the \textit{result} register, corresponding to \textit{N\_add.} in \refFigure{fig:pipeline}. 
As discussed in Section~\ref{sec:decomp_mix_unitary}, since the elements of matrix $H_1$ are either $1$ or $-1$, the multiplication with $H_1$’s column elements is implemented in hardware using simple sign inversion logic combined with parallel addition. After completing one elemental ansatz operation in the entire pipeline, the value in \textit{state} is re-initialized with the accumulated value in \textit{result}. The algorithm for this accumulation operation is as follows:

\scriptsize
\begin{verbatim}
   for(i=0; i<NUM_STATE; i++) {
     if(H_1[i][count_5th] == 1) result[i] += mult;
     else if(H_1[i][count_5th] == -1) result[i] -= mult;
     else printf("Error: Unexpected value in H_1");}
\end{verbatim}
\normalsize

%

\section{Evaluation}

\begin{figure}[t]
	\centering
	\includegraphics[width=0.5\textwidth]{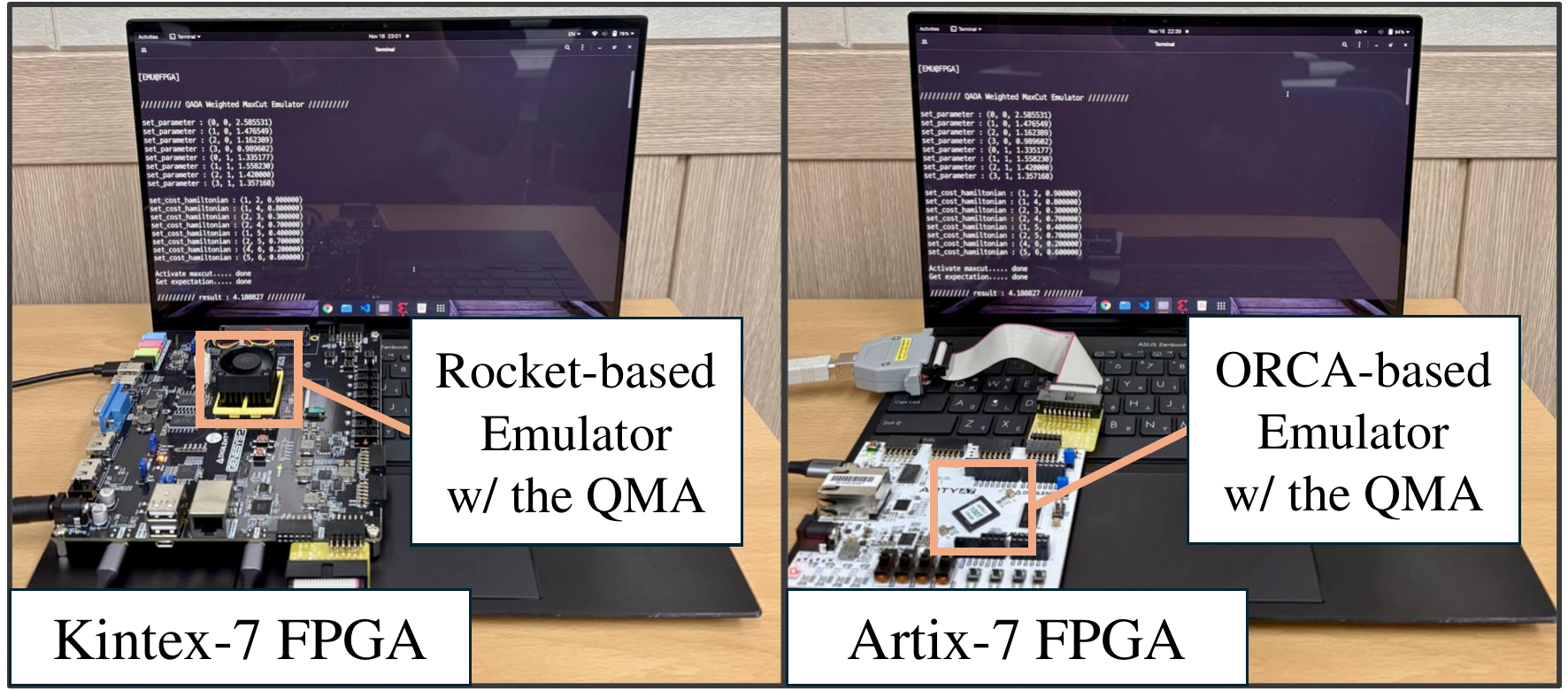}
	\caption{Execution of the prototyped QAOA emulator on the Kintex-7 FPGA board (left), and on Artix-7 FPGA board (right).}
	\label{fig:result_on_fpga}
\end{figure}

\begin{table}
\begin{center}
\caption{Comparison of execution time and energy consumption for 8-layer QAOA emulation with varying numbers of qubits between emulators with and without the QMA.}
		\centering
		\resizebox{\columnwidth}{!}{
		\begin{tabular}{ccccc}
			\toprule
			\multirow{2}{*}{\makecell{\\$n$}} & \multicolumn{2}{c}{w/o QMA} & \multicolumn{2}{c}{Proposed} \\
			\cmidrule(lr){2-3} \cmidrule(lr){4-5}
			& \multicolumn{1}{c}{Exec. time\hspace{0.15em}(\textit{ms})} & \multicolumn{1}{c}{Energy\hspace{0.15em}(\textit{mJ/exec.})}& \multicolumn{1}{c}{Exec. time\hspace{0.15em}(\textit{ms})} & \multicolumn{1}{c}{Energy\hspace{0.15em}(\textit{mJ/exec.})} \\
			\midrule
			2 &  0.41  	&0.23	& 0.26  	&0.15\\
			3 &  0.98  		&0.54	 & 0.26  	&0.15\\
			4 &  2.1 		&1.2	& 0.26 &0.15\\
			6 &  16  			&8.6	 & 0.27 & 0.17\\
			8 &  196  		&108	 & 0.30 & 0.30\\
			9 &  742 	&409	 & 0.34 &0.48\\
			\bottomrule
		\end{tabular}
		}
	\label{table:exec_time}	
\end{center}
\end{table}
\normalsize

Using the developed QEX, we implemented QAOA emulators with varying numbers of qubits, each incorporating the QMA shown in \refFigure{fig:architecture}.
These emulators were synthesized using the Xilinx Vivado tool~\cite{Vivado} and prototyped on a Xilinx Kintex-7 XC7K325T FPGA~\cite{Kintex7} board, as illustrated in \refFigure{fig:result_on_fpga}. 
After verifying stable operation, we evaluated the acceleration performance of the QMA by measuring the execution time of the QAOA with 8-layers on the emulator prototype. 
Energy consumption was calculated based on the measured execution time and power consumption data reported by Vivado. 
For comparison, a baseline processor with the same architecture as the emulator, but without the QMA, was also developed and prototyped on the same FPGA board. 
A C-based software implementation of the QAOA program was executed on the baseline processor to measure software-based emulation times. 
The results, summarized in \refTable{table:exec_time}, demonstrate substantial performance gains as the number of qubits increases. 
For 2-qubit QAOA emulation, the QMA achieved about 1.56$\times$ reduction in execution time and 1.53$\times$ improvement in energy efficiency. However, as the number of qubits increases, the benefits grow exponentially, culminating in a speedup of up to 2,182$\times$ and an energy efficiency improvement of 852$\times$ for 9-qubit emulation.

\begin{table}
\setlength{\tabcolsep}{5pt} 
\begin{center}
	\caption{Resource usage and utilization on Kintex-7 FPGA for conventional and proposed emulators with Rocket core across varying qubit counts.}
	\centering
	\begin{minipage}{0.5\textwidth}
		\centering
		\resizebox{1\columnwidth}{!}{
		\begin{tabular}{cccc}
			\toprule
			\multirow{3}{*}{\makecell{\\ $n$}} & \multicolumn{3}{c}{Emulators with a conventional HW accelerator}  \\
			\cmidrule(lr){2-4}
			& \multicolumn{1}{c}{LUT (Res. Util.)} & \multicolumn{1}{c}{\makecell{FF (Res. Util.)}} & \multicolumn{1}{c}{\makecell{DSP (Res. Util.)}} \\
			& \multicolumn{1}{c}{(Max : 203,800)} & \multicolumn{1}{c}{\makecell{(Max : 407,600)}} & \multicolumn{1}{c}{\makecell{(Max : 840)}} \\
			\midrule

			2 & \hspace{0.5em}43,086 (21.1\%)&  26,814 \hspace{0.5em}(6.6\%) & 103 (12.3\%)   \\
			3 & 136,437 (66.9\%) & 30,412 \hspace{0.5em}(7.5\%) & 315 (37.5\%) \\
			4\hspace{0.5em}(\textit{falied})& 1,275,722 (626\%)\hspace{1em}& 41,146 (10.1\%)& 834 (99.3\%)\\
			\bottomrule
		\end{tabular}
		}
	\end{minipage}
	
	
	\begin{minipage}{0.5\textwidth}
		\centering
		\resizebox{1\columnwidth}{!}{
			\begin{tabular}{cccc}
			\toprule
			\multirow{3}{*}{\makecell{\\ $n$}} & \multicolumn{3}{c}{Emulators with the proposed QMA}  \\
			\cmidrule(lr){2-4}
			& \multicolumn{1}{c}{LUT (Res. Util.)} & \multicolumn{1}{c}{\makecell{FF (Res. Util.)}} & \multicolumn{1}{c}{\makecell{DSP (Res. Util.)}} \\
& \multicolumn{1}{c}{(Max : 203,800)} & \multicolumn{1}{c}{\makecell{(Max : 407,600)}} & \multicolumn{1}{c}{\makecell{(Max : 840)}} \\
			\midrule

			2 & \hspace{0.5em}28,999 \hspace{0.5em}(14.2\%) 	& \hspace{0.5em}27,011 \hspace{0.5em}(6.6\%) 	& 31 (3.7\%) \\
			3 & \hspace{0.5em}29,486 \hspace{0.5em}(14.5\%) 	& \hspace{0.5em}27,524 \hspace{0.5em}(6.8\%) 	& 31 (3.7\%) \\
			4 & \hspace{0.5em}30,510 \hspace{0.5em}(15.0\%) 	& \hspace{0.5em}28,597 \hspace{0.5em}(7.0\%) 	& 31 (3.7\%) \\
			6 & \hspace{0.5em}36,369 \hspace{0.5em}(17.8\%) 	& \hspace{0.5em}34,813 \hspace{0.5em}(8.5\%) 	& 32 (3.8\%) \\
			8 & \hspace{0.5em}79,252 \hspace{0.5em}(38.9\%) 	& \hspace{0.5em}60,724 (14.9\%) 				& 32 (3.8\%) \\
			9 & 130,134 \hspace{0.5em}(63.9\%) 				& \hspace{0.5em}94,978 (23.3\%) 				& 32 (3.8\%) \\
			10\hspace{0.5em}(\textit{falied}) & 230,183 (113.0\%) 							& 163,157 (40.0\%) 							& 32 (3.8\%) \\
			\bottomrule
		\end{tabular}
		}
	\end{minipage}
	\label{table:fpga_resource_genesys2}
\end{center}
\scriptsize \centering * Ref., LUT/FF/DSP of the emulator w/o the QC accelerator : 24853 / 24445 / 4
\end{table}
\normalsize

\begin{table}
\begin{center}
	\caption{Resource usage and utilization on Artix-7 FPGA for proposed emulators with ORCA core across varying qubit counts.}
	\centering
		\centering
		\resizebox{1\columnwidth}{!}{
		\begin{tabular}{cccc}
			\toprule
			\multirow{3}{*}{\makecell{\\ $n$}} & \multicolumn{3}{c}{ORCA-based emulator employing the proposed QMA}  \\
			\cmidrule(lr){2-4}
			& \multicolumn{1}{c}{LUT (Res. Util.)} & \multicolumn{1}{c}{\makecell{FF (Res. Util.)}} & \multicolumn{1}{c}{\makecell{DSP (Res. Util.)}} \\
& \multicolumn{1}{c}{(Max : 63,400)} & \multicolumn{1}{c}{\makecell{(Max : 126,800)}} & \multicolumn{1}{c}{\makecell{(Max : 240)}} \\
			\midrule

			2 & 17,053 \hspace{0.5em}(26.9\%) & 19,785 (15.6\%) & 31 (12.9\%)   \\
			3 & 17,625 \hspace{0.5em}(27.8\%) & 20,402 (16.1\%) & 31 (12.9\%) \\
			4 & 18.748 \hspace{0.5em}(29.6\%) & 21,395 (16.9\%) & 31 (12.9\%) \\
			6 & 25,554 \hspace{0.5em}(40.3\%) & 27,806 (27.8\%) & 32 (13.3\%)\\
			8 & 51,168 \hspace{0.5em}(80.7\%) & 53,545 (42.2\%) & 32 (13.3\%)\\
			9\hspace{0.5em}(\textit{falied}) & 84,739 (133.7\%) & 87,692 (69.2\%) & 32 (13.3\%)\\
			\bottomrule
		\end{tabular}
		}
	\label{table:fpga_resource_arty}
\end{center}
\scriptsize \centering * Ref., LUT/FF of the emulator w/o the QC accelerator : 12555 / 17154 / 4
\end{table}
\normalsize


The reduction in execution time achieved through hardware parallelization inevitably increases hardware resource utilization. To address this, we proposed methods to mitigate the resource overhead effectively.  
To evaluate the effectiveness of these methods, we conducted experiments comparing hardware accelerators for matrix-vector multiplication in universal QC (\textit{Conventional}), designed without our optimizations, to the proposed QMA.  
These conventional accelerators, tailored to varying numbers of qubits, were integrated into the baseline platform shown in \refFigure{fig:architecture}, replacing the QMA. 
The RTL designs were adjusted for each qubit configuration.  
Both conventional and proposed emulators were prototyped on a Kintex-7 FPGA, incrementally increasing the number of qubits to compare resource utilization. The results are summarized in \refTable{table:fpga_resource_genesys2}.  
As shown, the conventional emulator became unsynthesizable beyond 3 qubits due to excessive resource usage. 
In contrast, the proposed emulator demonstrated efficient resource management, achieving a 78\% reduction in LUT utilization for the 3-qubit configuration and enabling seamless synthesis of the 3-qubit emulator, while supporting successful prototyping for configurations up to 9 qubits.

In addition, to evaluate the effectiveness of our proposed techniques on entry-level FPGA boards alongside mid-tier options, we conducted FPGA prototyping and resource usage comparisons for the proposed emulators integrated with the ORCA core~\cite{Vectorblox:ORCA} on the Xilinx Artix-7 XC7A100T board~\cite{Artix7} (cf. \refFigure{fig:result_on_fpga}). 
As shown in TABLE \ref{table:fpga_resource_arty}, the proposed emulator successfully scaled up to 8-qubits, whereas the conventional design was limited to a maximum of 2-qubits on the same board. 
These results demonstrate that the proposed emulator can efficiently support the emulation of more complex Weighted-MaxCut problems even in environments with severely constrained hardware resources.


\section{Conclusion}
We proposed an efficient approach to emulating QAOA in resource-constrained environments. 
Focusing on the Weighted-MaxCut problem, we optimized the conventional QAOA by restructuring its operations for hardware efficiency and designing a pipeline-based architecture for embedded systems. 
These efforts culminated in the development of the Quantum MaxCut Accelerator (QMA), which was integrated into custom emulators. 
To facilitate the rapid creation of QAOA emulators, we introduced QEX, a tool for efficient design and deployment. 
For evaluation, we implemented FPGA prototypes of both the proposed and baseline emulators, enabling comprehensive comparisons of performance, energy efficiency, and resource utilization.
Experimental results demonstrated the significant acceleration, energy savings, and resource efficiency achieved with the QMA and our approach.
These achievements provide a groundwork for advancing QAOA emulation in resource-constrained environments and enhancing QC capabilities.


\bibliographystyle{IEEEtran}
\bibliography{references_real}

\end{document}